\documentclass{article}
\usepackage{fullpage}
\usepackage{graphics,psfrag,epsfig}
\usepackage{amsfonts,latexsym,eucal,amsmath,amsthm,amssymb}

\begin{document}

\newcommand{\rum}{\rule{0.5pt}{0pt}}
\newcommand{\rub}{\rule{1pt}{0pt}}
\newcommand{\rim}{\rule{0.3pt}{0pt}}
\newcommand{\numtimes}{\mbox{\raisebox{1.5pt}{${\scriptscriptstyle \times}$}}}

\renewcommand{\refname}{References}

\twocolumn[%
\begin{center}
{\Large\bf Complexity Science for Simpletons\rule{0pt}{13pt}}\par
\bigskip
Craig Alan Feinstein \\
{\small\it 2712 Willow Glen Drive, Baltimore, Maryland 21209\rule{0pt}{13pt}}\\
\raisebox{-1pt}{\footnotesize E-mail: cafeinst@msn.com, BS"D}\par
\bigskip\smallskip
{\small\parbox{11cm}{%
In this article, we shall describe some of the most interesting
topics in the subject of Complexity Science for a general audience.
Anyone with a solid foundation in high school mathematics (with some
calculus) and an elementary understanding of computer programming
will be able to follow this article. First, we shall explain the
significance of the $P$ versus $NP$ problem and solve it. Next, we
shall describe two other famous mathematics problems, the Collatz
$3n+1$ Conjecture and the Riemann Hypothesis, and show how both
Chaitin's incompleteness theorem and Wolfram's notion of
``computational irreducibility" are important for understanding why
no one has, as of yet, solved these two
problems.

\bigskip \noindent \textbf{Disclaimer:} This article was authored
by Craig Alan Feinstein in his private capacity. No official
support or endorsement by the U.S. Government is intended or
should be inferred.\rule[0pt]{0pt}{0pt}}}\bigskip
\end{center}]{%

\section{Challenge}

Imagine that you have a collection of one billion lottery tickets
scattered throughout your basement in no particular order. An
official from the lottery announces the number of the winning
lottery ticket. For a possible prize of one billion dollars, is it a
good idea to search your basement until you find the winning ticket
or until you come to the conclusion that you do not possess the
winning ticket? Most people would think not - even if the winning
lottery ticket were in your basement, performing such a search could
take $10^9/(60 \times 60 \times 24 \times 365.25)$ years, over
thirty work-years, assuming that it takes you at least one second to
examine each lottery ticket. Now imagine that you have a collection
of only one thousand lottery tickets in your basement. Is it a good
idea to search your basement until you find the winning ticket or
until you come to the conclusion that you do not possess the winning
ticket? Most people would think so, since doing such would take at
most a few hours.

From these scenarios, let us postulate a general rule that the
maximum time that it may take for one person to search $N$ unsorted
objects for one specific object is directly proportional to $N$.
This is clearly the case for physical objects, but what about
abstract objects? For instance, let us suppose that a dating service
is trying to help $n$ single women and $n$ single men to get
married. Each woman gives the dating service a list of
characteristics that she would like to see in her potential husband,
for instance, handsome, caring, athletic, domesticated, etc. And
each man gives the dating service a list of characteristics that he
would like to see in his potential wife, for instance, beautiful,
obedient, good cook, thrifty, etc. The dating service is faced with
the task of arranging dates for each of its clients so as to satisfy
everyone's preferences.

Now there are $n!$ (which is shorthand for $n \times (n-1) \times
(n-2) \times...\times 2 \times 1$) possible ways for the dating
service to arrange dates for each of its clients, but only a
fraction of such arrangements would satisfy all of its clients. If
$n=100$, it would take too long for the dating service's computer to
evaluate all $100!$ possible arrangements until it finds an
arrangement that would satisfy all of its clients. ($100!$ is too
large a number of possibilities for any modern computer to handle.) Is
there an efficient way for the dating service's computer to find
dates with compatible potential spouses for each of the dating
service's clients so that everyone is happy, assuming that it is
possible to do such? Yes, and here is how:

\bigskip\noindent \textbf{Matchmaker Algorithm -} Initialize the set
$M=\emptyset$. Search for a list of compatible relationships between
men and women that alternates between a compatible relationship
$\{x_1,x_2\}$ not contained in set $M$, followed by a compatible
relationship $\{x_2,x_3\}$ contained in set $M$, followed by a
compatible relationship $\{x_3,x_4\}$ not contained in set $M$,
followed by a compatible relationship $\{x_4,x_5\}$ contained in set
$M$, and so on, ending with a compatible relationship
$\{x_{m-1},x_m\}$ not contained in set $M$, where both $x_1$ and
$x_m$ are not members of any compatible relationships contained in
set $M$. Once such a list is found, for each compatible relationship
$\{x_i,x_{i+1}\}$ in the list, add $\{x_i,x_{i+1}\}$ to $M$ if
$\{x_i,x_{i+1}\}$ is not contained in $M$ or remove
$\{x_i,x_{i+1}\}$ from $M$ if $\{x_i,x_{i+1}\}$ is contained in $M$.
(Note that this procedure must increase the size of set $M$ by one.)
Repeat this procedure until no such list exists.

\bigskip Such an algorithm is guaranteed to \underline{efficiently}
find an arrangement $M$ that will satisfy all of the dating
service's clients whenever such an arrangement exists \cite{b:PS82}.
So we see that with regard to abstract objects, it is not
necessarily the case that the maximum time that it may take for one
to search $N$ unsorted objects for a specific object is directly
proportional to $N$; in the dating service example, there are $n!$
possible arrangements between men and women, yet it is not necessary
for a computer to examine all $n!$ arrangements in order to find a
satisfactory arrangement. One might think that the problem of
finding a satisfactory dating arrangement is easy for a modern
computer to solve because the list of pairs of men and women who are
compatible is relatively small (of size at most $n^2$, which is much
smaller than the number of possible arrangements $n!$) and because
it is easy to verify whether any particular arrangement will make
everyone happy. But this reasoning is invalid, as we shall
demonstrate:

\section{The SUBSET-SUM Problem}

Consider the following problem: You are given a set
$A=\{a_1,...,a_n\}$ of $n$ integers and another integer $b$ which we
shall call the \textit{target integer}. You want to know if there
exists a subset of $A$ for which the sum of its elements is equal to
$b$. (We shall consider the sum of the elements of the empty set to
be zero.) This problem is called the \textit{SUBSET-SUM problem}
\cite{b:CLR90}. Now, there are $2^n$ subsets of $A$, so one could
naively solve this problem by exhaustively comparing the sum of the
elements of each subset of $A$ to $b$ until one finds a subset-sum
equal to $b$, but such a procedure would be infeasible for even the
fastest computers in the world to implement when $n=100$. Is there
an algorithm which can considerably reduce the amount of work for
solving the SUBSET-SUM problem? Yes, there is an algorithm
discovered by Horowitz and Sahni in 1974 \cite{b:HS74}, which we
shall call the \textit{Meet-in-the-Middle algorithm}, that takes on
the order of $2^{n/2}$ steps to solve the SUBSET-SUM problem instead
of the $2^n$ steps of the naive exhaustive comparison algorithm:

\bigskip\noindent \textbf{Meet-in-the-Middle Algorithm -}
First, partition the set $A$ into two subsets, $A^+
=\{a_1,...,a_{\lceil \frac{n}{2} \rceil}\}$ and $A^- =\{a_{\lceil
\frac{n}{2} \rceil+1},...,a_n\}$. Let us define $S^+$ and $S^-$ as
the sets of subset-sums of $A^+$ and $A^-$, respectively. Sort sets
$S^+$ and $b-S^-$ in ascending order. Compare the first elements in
both of the lists. If they match, then stop and output that there is
a solution. If not, then compare the greater element with the next
element in the other list. Continue this process until there is a
match, in which case there is a solution, or until one of the lists
runs out of elements, in which case there is no solution.

\bigskip This algorithm takes on the order of $2^{n/2}$ steps, since
it takes on the order of $2^{n/2}$ steps to sort sets $S^+$ and
$b-S^-$ (assuming that the computer can sort in linear-time) and on
the order of $2^{n/2}$ steps to compare elements from the sorted
lists $S^+$ and $b-S^-$. Are there any faster algorithms for solving
SUBSET-SUM? $2^{n/2}$ is still a very large number when $n=100$,
even though this strategy is a vast improvement over the naive
strategy. It turns out that no algorithm with a better worst-case
running-time has ever been found since the Horowitz and Sahni paper
\cite{b:Woe03}. And the reason for this is because it is impossible
for such an algorithm to exist. Here is an explanation why:

\bigskip \noindent \textit{Explanation:} To understand why there
is no algorithm with a faster worst-case running-time than the
Meet-in-the-Middle algorithm, let us travel back in time
seventy-five years, long before the internet. If one were to ask
someone back then what a computer is, one would have gotten the
answer, ``a person who computes (usually a woman)" instead of the
present day definition, ``a machine that computes" \cite{b:Gri05}.
Let us imagine that we knew two computers back then named Mabel and
Mildred (two popular names for women in the 1930's \cite{b:Sha98}).
Mabel is very efficient at sorting lists of integers into ascending
order; for instance she can sort a set of ten integers in 15
seconds, whereas it takes Mildred 20 seconds to perform the same
task. However, Mildred is very efficient at comparing two integers
$a$ and $b$ to determine whether $a<b$ or $a=b$ or $a>b$; she can
compare ten pairs of integers in 15 seconds, whereas it takes Mabel
20 seconds to perform the same task.

Let's say we were to give both Mabel and Mildred the task of
determining whether there exists a subset of some four element set,
$A=\{a_1,a_2,a_3,a_4\}$, for which the sum of its elements adds up
to $b$. Since Mildred is good at comparing but not so good at
sorting, Mildred chooses to solve this problem by comparing $b$ to
all of the sixteen subset-sums of $A$. Since Mabel is good at
sorting but not so good at comparing, Mabel decides to solve this
problem by using the Meet-in-the-Middle algorithm. In fact, of all
algorithms that Mabel could have chosen to solve this problem, the
Meet-in-the-Middle algorithm is the most efficient for her to use on
sets $A$ with only four integers. And of all algorithms that Mildred
could have chosen to solve this problem, comparing $b$ to all of the
sixteen subset-sums of $A$ is the most efficient algorithm for her
to use on sets $A$ with only four integers.

Now we are going to use the principle of mathematical induction to
prove that the best algorithm for Mabel to use for solving the
SUBSET-SUM problem for large $n$ is the Meet-in-the-Middle
algorithm: We already know that this is true when $n=4$. Let us
assume that this is true for $n$, i.e., that of all possible
algorithms for Mabel to use for solving the SUBSET-SUM problem on
sets with $n$ integers, the Meet-in-the-Middle algorithm has the
best worst-case running-time. Then we shall prove that this is also
true for $n+1$:

Let $S$ be the set of all subset-sums of the set
$A=\{a_1,a_2,...,a_n\}$. Notice that the SUBSET-SUM problem on the
set $A \cup \{a'\}$ of $n+1$ integers and target $b$ is equivalent
to the problem of determining whether (1) $b \in S$ or (2) $b' \in
S$ (where $b'=b-a'$). (The symbol $\in$ means ``is a member of".)
Also notice that these two subproblems, (1) and (2), are independent
from one another in the sense that the values of $b$ and $b'$ are
unrelated to each other and are also unrelated to set $S$;
therefore, in order to determine whether $b \in S$ or $b' \in S$, it
is necessary to solve both subproblems (assuming that the first
subproblem solved has no solution). So it is clear that if Mabel
could solve both subproblems in the fastest time possible and also
whenever possible make use of information obtained from solving
subproblem (1) to save time solving subproblem (2) and whenever
possible make use of information obtained from solving subproblem
(2) to save time solving subproblem (1), then Mabel would be able to
solve the problem of determining whether (1) $b \in S$ or (2) $b'
\in S$ in the fastest time possible.

We shall now explain why the Meet-in-the-Middle algorithm has this
characteristic for sets of size $n+1$: It is clear that by the
induction hypothesis, the Meet-in-the-Middle algorithm solves each
subproblem in the fastest time possible, since it works by applying
the Meet-in-the-Middle algorithm to each subproblem, without loss of
generality sorting and comparing elements in lists $S^+$ and $b-S^-$
and also sorting and comparing elements in lists $S^+$ and $b'-S^-$
as the algorithm sorts and compares elements in lists $S^+$ and
$b-[S^- \cup (S^- + a')]$. There are two situations in which it is
possible for the Meet-in-the-Middle algorithm to make use of
information obtained from solving subproblem (1) to save time
solving subproblem (2) or to make use of information obtained from
solving subproblem (2) to save time solving subproblem (1). And the
Meet-in-the-Middle algorithm takes advantage of both of these
opportunities:

\begin{itemize}
\item Whenever the Meet-in-the-Middle algorithm compares two elements
from lists $S^+$ and $b-S^-$ and the element in list $S^+$ turns out
to be less than the element in list $b-S^-$, the algorithm makes use
of information obtained from solving subproblem (1) (the fact that
the element in list $S^+$ is less than the element in list $b-S^-$)
to save time, when $n$ is odd, solving subproblem (2) (the algorithm
does not consider the element in list $S^+$ again).
\item Whenever the Meet-in-the-Middle algorithm compares two elements
from lists $S^+$ and $b'-S^-$ and the element in list $S^+$ turns
out to be less than the element in list $b'-S^-$, the algorithm
makes use of information obtained from solving subproblem (2) (the
fact that the element in list $S^+$ is less than the element in list
$b'-S^-$) to save time, when $n$ is odd, solving subproblem (1) (the
algorithm does not consider the element in list $S^+$ again).
\end{itemize}

\noindent Therefore, we can conclude that the Meet-in-the-Middle
algorithm whenever possible makes use of information obtained from
solving subproblem (1) to save time solving subproblem (2) and
whenever possible makes use of information obtained from solving
subproblem (2) to save time solving subproblem (1). So we have
completed our induction step to prove true for $n+1$, assuming true
for $n$.

Therefore, the best algorithm for Mabel to use for solving the
SUBSET-SUM problem for large $n$ is the Meet-in-the-Middle
algorithm. But is the Meet-in-the-Middle algorithm the best
algorithm for Mildred to use for solving the SUBSET-SUM problem for
large $n$? Since the Meet-in-the-Middle algorithm is not the fastest
algorithm for Mildred to use for small $n$, is it not possible that
the Meet-in-the-Middle algorithm is also not the fastest algorithm
for Mildred to use for large $n$? It turns out that for large $n$,
there is no algorithm for Mildred to use for solving the SUBSET-SUM
problem with a faster worst-case running-time than the
Meet-in-the-Middle algorithm. Why?

Notice that the Meet-in-the-Middle algorithm takes on the order of
$2^{n/2}$ steps regardless of whether Mabel or Mildred applies it.
And notice that the algorithm of naively comparing the target $b$ to
all of the $2^n$ subset-sums of set $A$ takes on the order of $2^n$
steps regardless of whether Mabel or Mildred applies it. So for
large $n$, regardless of who the computer is, the Meet-in-the-Middle
algorithm is faster than the naive exhaustive comparison algorithm -
from this example, we can understand the general principle that the
asymptotic running-time of an algorithm does not differ by more than
a polynomial factor when run on different types of computers
\cite{b:Woe03, b:Wol85}. Therefore, since no algorithm is faster
than the Meet-in-the-Middle algorithm for solving SUBSET-SUM for
large $n$ when applied by Mabel, we can conclude that no algorithm
is faster than the Meet-in-the-Middle algorithm for solving
SUBSET-SUM for large $n$ when applied by Mildred. And furthermore,
using this same reasoning, we can conclude that no algorithm is
faster than the Meet-in-the-Middle algorithm for solving SUBSET-SUM
for large $n$ when run on a modern computing machine. \qed

\bigskip So it doesn't matter whether the computer is Mabel, Mildred,
or any modern computing machine; the fastest algorithm which solves
the SUBSET-SUM problem for large $n$ is the Meet-in-the-Middle
algorithm. Because once a solution to the SUBSET-SUM problem is
found, it is easy to verify (in polynomial-time) that it is indeed a
solution, we say that the SUBSET-SUM problem is in class $NP$
\cite{b:BC94}. And because there is no algorithm which solves
SUBSET-SUM that runs in polynomial-time (since the
Meet-in-the-Middle algorithm runs in exponential-time and is the
fastest algorithm for solving SUBSET-SUM, as we have shown above),
we say that the SUBSET-SUM problem is not in class $P$
\cite{b:BC94}. Then since the SUBSET-SUM problem is in class $NP$
but not in class $P$, we can conclude that $P \neq NP$, thus solving
the $P$ versus $NP$ problem. The solution to the $P$
versus $NP$ problem demonstrates that it is possible to hide
abstract objects (in this case, a subset of set $A$) without an
abundance of resources - it is, in general, more difficult to find a
subset of a set of only one hundred integers for which the sum of
its elements equals a target integer than to find the winning
lottery-ticket in a pile of one billion unsorted lottery tickets,
even though the lottery-ticket problem requires much more resources
(one billion lottery tickets) than the SUBSET-SUM problem requires
(a list of one hundred integers).

\section{Does $P \neq NP$ really matter?}

Even though $P \neq NP$, might there still be algorithms which
efficiently solve problems that are in $NP$ but not $P$ in the
average-case scenario? (Since the $P \neq NP$ result deals only with
the worst-case scenario, there is nothing to forbid this from
happening.) The answer is yes; for many problems which are in $NP$
but not $P$, there exist algorithms which efficiently solve them in
the average-case scenario \cite{b:Odl90, b:Wil84}, so the statement
that $P \neq NP$ is not as ominous as it sounds. In fact, there is a
very clever algorithm which solves almost all instances of the
SUBSET-SUM problem in polynomial-time \cite{b:CJL92, b:MvOV96,
b:Odl90}. (The algorithm works by converting the SUBSET-SUM problem
into the problem of finding the shortest non-zero vector of a
lattice, given its basis.) But even though for many problems which
are in $NP$ but not $P$, there exist algorithms which efficiently
solve them in the average-case scenario, in the opinion of most
complexity-theorists, it is probably false that for all problems
which are in $NP$ but not $P$, there exist algorithms which
efficiently solve them in the average-case scenario \cite{b:BCGL92}.

Even though $P \neq NP$, might it still be possible that there exist
polynomial-time \textit{randomized} algorithms which correctly solve
problems in $NP$ but not in $P$ with a high probability regardless
of the problem instance? (The word ``randomized" in this context
means that the algorithm bases some of its decisions upon random
variables. The advantage of these types of algorithms is that
whenever they fail to output a solution, there is still a good
chance that they will succeed if they are run again.) The answer is
probably no, as there is a widely believed conjecture that $P=BPP$,
where $BPP$ is the class of decision problems for which there are
polynomial-time randomized algorithms that correctly solve them at
least two-thirds of the time regardless of the problem instance
\cite{b:IW97}.

\section{Are Quantum Computers the Answer?}

A \textit{quantum} \textit{computer} is any computing device which
makes direct use of distinctively quantum mechanical phenomena, such
as superposition and entanglement, to perform operations on data. As
of today, the field of practical quantum computing is still in its
infancy; however, much is known about the theoretical properties of
a quantum computer. For instance, quantum computers have been shown
to efficiently solve certain types of problems, like factoring
integers \cite{b:Sho94}, which are believed to be difficult to solve
on a \textit{classical computer}, e.g., a human-computer like Mabel
or Mildred or a machine-computer like an IBM PC or an Apple
Macintosh.

Is it possible that one day quantum computers will be built and will
solve problems like the SUBSET-SUM problem efficiently in
polynomial-time? The answer is that it is generally suspected by
complexity theorists to be impossible for a quantum computer to
solve the SUBSET-SUM problem (and all other problems which share a
characteristic with the SUBSET-SUM problem in that they belong to a
subclass of $NP$ problems known as \textit{NP-complete} problems
\cite{b:BC94}) in polynomial-time. A curious fact is that if one
were to solve the SUBSET-SUM problem on a quantum computer by brute
force, the algorithm would have a running-time on the order of
$2^{n/2}$ steps, which (by coincidence?) is the same asymptotic
running-time of the fastest algorithm which solves SUBSET-SUM on a
classical computer, the Meet-in-the-Middle algorithm \cite{b:Aar05,
b:BBBV97, b:Gro96}.

In any case, no one has ever built a practical quantum computer and
some scientists are even of the opinion that building such a
computer is impossible; the acclaimed complexity theorist, Leonid
Levin, wrote: ``QC of the sort that factors long numbers seems
firmly rooted in science fiction. It is a pity that popular accounts
do not distinguish it from much more believable ideas, like Quantum
Cryptography, Quantum Communications, and the sort of Quantum
Computing that deals primarily with locality restrictions, such as
fast search of long arrays. It is worth noting that the reasons why
QC must fail are by no means clear; they merit thorough
investigation. The answer may bring much greater benefits to the
understanding of basic physical concepts than any factoring device
could ever promise. The present attitude is analogous to, say,
Maxwell selling the Daemon of his famous thought experiment as a
path to cheaper electricity from heat. If he did, much of insights
of today's thermodynamics might be lost or delayed \cite{b:Lev03}."

\section{Unprovable Conjectures}

In the early twentieth century, the famous mathematician, David
Hilbert, proposed the idea that all mathematical facts can be
derived from only a handful of self-evident axioms. In the 1930's,
Kurt G\"{o}del proved that such a scenario is impossible by showing
that for any proposed finite axiom system for arithmetic, there must
always exist true statements that are unprovable within the
system, if one is to assume that the axiom system has no
inconsistencies. Alan Turing extended this result to show that it is
impossible to design a computer program which can determine whether
any other computer program will eventually halt. In the latter half
of the 20th century, Gregory Chaitin defined a real number between
zero and one, which he calls $\Omega$, to be
the probability that a computer program halts. And Chaitin proved that:

\bigskip \noindent \textbf{Theorem 1 -} For any mathematics problem,
the bits of $\Omega$, when $\Omega$ is expressed in binary,
completely determine whether that problem is solvable or not.

\bigskip \noindent \textbf{Theorem 2 -} The bits of $\Omega$ are random
and only a finite number of them are even possible to know.

\bigskip \noindent From these two theorems, it follows that the very
structure of mathematics itself is random and mostly unknowable! \cite{b:Cha05}

Even though Hilbert's dream to be able derive every mathematical
fact from only a handful of self-evident axioms was destroyed by
G\"{o}del in the 1930's, this idea has still had an enormous impact
on current mathematics research \cite{b:Zac05}. In fact, even though
mathematicians as of today accept the incompleteness theorems proven
by G\"{o}del, Turing, and Chaitin as true, in general these same
mathematicians also believe that these incompleteness theorems
ultimately have no impact on traditional mathematics research, and
they have thus adopted Hilbert's paradigm of deriving mathematical
facts from only a handful of self-evident axioms as a practical way
of researching mathematics. Gregory Chaitin has been warning these
mathematicians for decades now that these incompleteness theorems
are actually very relevant to advanced mathematics research, but the
overwhelming majority of mathematicians have not taken
his warnings seriously \cite{b:Cha03}. We shall directly confirm
Chaitin's assertion that incompleteness is indeed very relevant to
advanced mathematics research by giving very strong evidence that
two famous mathematics problems, determining whether the Collatz
$3n+1$ Conjecture is true and determining whether the Riemann
Hypothesis is true, are impossible to solve:

\bigskip\noindent \textbf{The Collatz $3n+1$ Conjecture -} Here's a
fun experiment that you, the reader, can try: Pick any positive
integer, $n$. If $n$ is even, then compute $n/2$ or if $n$ is odd,
then compute $(3n+1)/2$. Then let $n$ equal the result of this
computation and perform the whole procedure again until $n=1$. For
instance, if you had chosen $n=11$, you would have obtained the
sequence $(3 \times 11 \mbox{ } + \mbox{ } 1)/2=17$, $(3 \times 17
\mbox{ } + \mbox{ } 1)/2=26$, $26/2=13$, 20, 10, 5, 8, 4, 2, 1.

The Collatz $3n+1$ Conjecture states that such an algorithm will
always eventually reach $n=1$ and halt \cite{b:Lag85}. Computers
have verified this conjecture to be true for all positive integers
less than $224 \times 2^{50} \approx$ $2.52 \times 10^{17}$
\cite{b:Roo03}. Why does this happen? One can give an informal
argument as to why this may happen \cite{b:Cra78} as follows: Let us
assume that at each step, the probability that $n$ is even is
one-half and the probability that $n$ is odd is one-half. Then at
each iteration, $n$ will decrease by a multiplicative factor of
about $(\frac{3}{2})^{1/2}(\frac{1}{2})^{1/2}=$
$(\frac{3}{4})^{1/2}$ on average, which is less than one; therefore,
$n$ will eventually converge to one with probability one. But such
an argument is \underline{not} a rigorous mathematical proof, since
the probability assumptions that the argument is based upon are not
well-defined and even if they were well-defined, it would still be
possible (although extremely unlikely, with probability zero) that
the algorithm will never halt for some input.

Is there a rigorous mathematical proof of the Collatz $3n+1$
Conjecture? As of today, no one has found a rigorous proof that the
conjecture is true and no one has found a rigorous proof that the
conjecture is false. In fact, Paul Erd\"{o}s, who was one of the
greatest mathematicians of the twentieth century, commented about
the Collatz $3n+1$ Conjecture: ``Mathematics is not yet ready for
such problems \cite{b:Lag85}." We can informally demonstrate that
there is no way to deductively prove that the conjecture is true, as
follows:

\bigskip \noindent \textit{Explanation:} First, notice that in order
to be certain that the algorithm will halt for a given input $n$, it
is necessary to know whether the value of $n$ at the beginning of
each iteration of the algorithm is even or odd. (For a rigorous
proof of this, see ``The Collatz Conjecture is Unprovable"
\cite{b:Fei05}.) For instance, if the algorithm starts with input
$n=11$, then in order to know that the algorithm halts at one, it is
necessary to know that 11 is odd, $(3 \times 11 \mbox{ } + \mbox{ }
1)/2=17$ is odd, $(3 \times 17 \mbox{ } + \mbox{ } 1)/2=26$ is even,
$26/2=13$ is odd, 20 is even, 10 is even, 5 is odd, 8 is even, 4 is
even, and 2 is even. We can express this information (odd, odd,
even, odd, even, even, odd, even, even, even) as a vector of zeroes
and ones, $(1,1,0,1,0,0,1,0,0,0)$. Let us call this vector the
\textit{parity vector} of $n$. (If $n$ never converges to one, then
its parity vector must be infinite-dimensional.) If one does not
know the parity vector of the input, then it is impossible to know
what the algorithm does at each iteration and therefore impossible
to be certain that the algorithm will converge to one. So any proof
that the algorithm applied to $n$ halts must specify the parity
vector of $n$. Next, let us give a definition of a \textit{random
vector}:

\bigskip\noindent \textbf{Definition -} We shall say that a vector
${\rm {\bf x}} \in \{0,1\}^m$ is \textit{random} if ${\rm {\bf x}}$
cannot be specified in less than $m$ bits in a computer text-file
\cite{b:Cha90}.

\bigskip\noindent\textbf{Example 1 -} The vector of one million
concatenations of the vector $(0,1)$ is \underline{not} random,
since we can specify it in less than two million bits in a computer
text-file. (We just did.)

\bigskip\noindent\textbf{Example 2 -} The vector of outcomes of one
million coin-tosses has a good chance of fitting our definition of
``random", since much of the time the most compact way of specifying
such a vector is to simply make a list of the results of each
coin-toss, in which one million bits are necessary.

\bigskip Now let us suppose that it were possible to prove the
Collatz $3n+1$ Conjecture and let $B$ be the number of bits in a
hypothetical computer text-file containing such a proof. And let
$(x_0,x_1,x_2,...,x_B)$ be a random vector, as defined above. (It is
not difficult to prove that at least half of all vectors with $B+1$
zeroes and ones are random \cite{b:Cha90}.) There is a mathematical
theorem \cite{b:Lag85} which says that there must exist a number $n$
with the first $B+1$ bits of its parity vector equal to
$(x_0,x_1,x_2,...,x_B)$; therefore, any proof of the Collatz $3n+1$
Conjecture must specify vector $(x_0,x_1,x_2,...,x_B)$ (as we
discussed above), since such a proof must show that the Collatz
algorithm halts when given input $n$. But since vector
$(x_0,x_1,x_2,...,x_B)$ is random, $B+1$ bits are required to
specify vector $(x_0,x_1,x_2,...,x_B)$, contradicting our assumption
that $B$ is the number of bits in a computer text-file containing a
proof of the Collatz $3n+1$ Conjecture; therefore, a formal proof of
the Collatz $3n+1$ Conjecture cannot exist \cite{b:Fei05}. \qed

\bigskip\noindent \textbf{The Riemann Hypothesis -} There is also
another famous unresolved conjecture, the Riemann Hypothesis, which
has a characteristic similar to that of the Collatz $3n+1$
Conjecture, in that it too can never be proven true. In the opinion
of many mathematicians, the Riemann Hypothesis is the most important
unsolved problem in mathematics \cite{b:Der03}. The reason why it is
so important is because a resolution of the Riemann Hypothesis would
shed much light on the distribution of prime numbers: It is well
known that the number of prime numbers less than $n$ is
approximately $\int_2^n \frac{dx}{\log x}$. If the Riemann
Hypothesis is true, then for large $n$, the error in this
approximation must be bounded by $c n^{1/2} \log n$ for some
constant $c>0$ \cite{b:Wei05b}, which is also a bound for a
\textit{random walk}, i.e., the sum of $n$ independent random
variables, $X_k$, for $k=1,2,...,n$ in which the probability that
$X_k=-c$ is one-half and the probability that $X_k=c$ is one-half.

The Riemann-Zeta function $\zeta(s)$ is a complex function which is
defined to be $\zeta(s)=\frac{s}{s-1}-s \int_1^{\infty}
\frac{x-\lfloor x \rfloor}{x^{s+1}} dx$ when the real part of the
complex number $s$ is positive. The Riemann Hypothesis states that
if $\rho=\sigma+ti$ is a complex root of $\zeta$ and $0<\sigma<1$,
then $\sigma=1/2$. It is well known that there are infinitely many
roots of $\zeta$ that have $0<\sigma<1$. And just like the Collatz
$3n+1$ Conjecture, the Riemann Hypothesis has been verified by
high-speed computers - for all $|t| < T$ where $T \approx 5 \times
10^8$ \cite{b:Odl89}. But it is still unknown whether there
exists a $|t| \geq T$ such that $\zeta(\sigma+ti)=0$, where $\sigma
\neq 1/2$. And just like the Collatz $3n+1$ Conjecture, one can give
a heuristic probabilistic argument that the Riemann Hypothesis is
true \cite{b:GC68}, as follows:

It is well known that the Riemann Hypothesis follows from the
assertion that for large $n$, $M(n)=\Sigma_{k=1}^n \mu(k)$ is
bounded by $c n^{1/2} \log n$ for some constant $c>0$, where $\mu$
is the M\"{o}bius Inversion function defined on $\mathbb N$ in which
$\mu(k)=-1$ if $k$ is the product of an odd number of distinct
primes, $\mu(k)=1$ if $k$ is the product of an even number of
distinct primes, and $\mu(k)=0$ otherwise. If we were to assume that
$M(n)$ is distributed as a random walk, which is certainly plausible
since there is no apparent reason why it should not be distributed
as a random walk, then by probability theory, $M(n)$ is bounded for
large $n$ by $c n^{1/2} \log n$ for some constant $c>0$, with
probability one; therefore, it is very likely that the Riemann
Hypothesis is true. We shall now explain why the Riemann Hypothesis
is unprovable, just like the Collatz $3n+1$ Conjecture:

\bigskip \noindent \textit{Explanation:} The Riemann Hypothesis is
equivalent to the assertion that for each $T>0$, the number of real
roots $t$ of $\zeta(1/2 + ti)$, where $0<t<T$, is equal to the
number of roots of $\zeta(s)$ in $\{s=\sigma+ti:\mbox{ }
0<\sigma<1,\mbox{ } 0<t<T\}$. It is well known that there exists a
continuous \underline{real} function $Z(t)$ (called the
Riemann-Siegel function) such that $|Z(t)| = |\zeta(1/2 + ti)|$, so
the real roots $t$ of $\zeta(1/2 + ti)$ are the same as the real
roots $t$ of $Z(t)$. (The formula for $Z(t)$ is $\zeta(1/2 + ti)e^{i
\vartheta(t)}$, where $\vartheta(t)= \arg [\Gamma ( \frac{1}{4} +
\frac{1}{2}it) ] - \frac{1}{2} t \ln \pi$.) Then because the formula
for the real roots $t$ of $\zeta(1/2 + ti)$ cannot be reduced to a
formula that is simpler than the equation, $\zeta(1/2 + ti)=0$, the
only way to determine the number of real roots $t$ of $\zeta(1/2 +
ti)$ in which $0<t<T$ is to count the changes in sign of the real
function $Z(t)$, where $0<t<T$ \cite{b:Pug98}.

So in order to prove that the number of real roots $t$ of $\zeta(1/2
+ ti)$, where $0<t<T$, is equal to the number of roots of $\zeta(s)$
in $\{s=\sigma+ti:\mbox{ } 0<\sigma<1,\mbox{ } 0<t<T\}$, which can
be computed via a theorem known as the \textit{Argument Principle}
\underline{without} counting the changes in sign of $Z(t)$, where
$0<t<T$ \cite{b:Odl94, b:Pug98, b:RS91}, it is necessary to count
the changes in sign of $Z(t)$, where $0<t<T$. (Otherwise, it would
be possible to determine the number of real roots $t$ of $\zeta(1/2
+ ti)$, where $0<t<T$, without counting the changes in sign of
$Z(t)$ by computing the number of roots of $\zeta(s)$ in
$\{s=\sigma+ti:\mbox{ } 0<\sigma<1,\mbox{ } 0<t<T\}$ via the
Argument Principle.) As $T$ becomes arbitrarily large, the time that
it takes to count the changes in sign of $Z(t)$, where $0<t<T$,
approaches infinity for the following reasons: (1) There are
infinitely many changes in sign of $Z(t)$. (2) The time that it
takes to evaluate the sign of $Z(t)$ approaches infinity as $t
\rightarrow \infty$ \cite{b:Pug98}. Hence, an infinite amount of
time is required to prove that for each $T>0$, the number of real
roots $t$ of $\zeta(1/2 + ti)$, where $0<t<T$, is equal to the
number of roots of $\zeta(s)$ in $\{s=\sigma+ti:\mbox{ }
0<\sigma<1,\mbox{ } 0<t<T\}$ (which is equivalent to proving the
Riemann Hypothesis), so the Riemann Hypothesis is unprovable.
\qed

\bigskip Chaitin's incompleteness theorem implies that mathematics
is filled with facts which are both true and unprovable, since it
states that the bits of $\Omega$ completely determine whether any
given mathematics problem is solvable and only a finite number of
bits of $\Omega$ are even knowable \cite{b:Cha05}. And we have shown
that there is a very good chance that both the Collatz $3n+1$
Conjecture and the Riemann Hypothesis are examples of such facts. Of
course, we can never formally prove that either one of these
conjectures is both true and unprovable, for obvious reasons. The
best we can do is prove that they are unprovable and provide
computational evidence and heuristic probabilistic reasoning to
explain why these two conjectures are most likely true, as we have
done. And of course, it is conceivable that one could find a
counter-example to the Collatz $3n+1$ Conjecture by finding a number
$n$ for which the Collatz algorithm gets stuck in a nontrivial cycle
or a counter-example to the Riemann Hypothesis by finding a complex
root, $\rho=\sigma+ti$, of $\zeta$ for which $0<\sigma<1$ and
$\sigma \neq 1/2$. But so far, no one has presented any such
counter-examples.

The theorems that the Collatz $3n+1$ Conjecture and the Riemann
Hypothesis are unprovable illustrate a point which Chaitin has been
making for years, that mathematics is not so much different from
empirical sciences like physics \cite{b:Cha05, b:Dom05}. For
instance, scientists universally accept the law of gravity to be
true based on experimental evidence, but such a law is by no means
absolutely certain - even though the law of gravity has been
observed to hold in the past, it is not inconceivable that the law
of gravity may cease to hold in the future. So too, in mathematics
there are conjectures like the Collatz $3n+1$ Conjecture and the
Riemann Hypothesis which are strongly supported by experimental
evidence but can never be proven true with absolute certainty.

\section{Computational Irreducibility}

\noindent Up until the last decade of the twentieth century, the
most famous unsolved problem in all of mathematics was to prove the
following conjecture:

\bigskip\noindent \textbf{Fermat's Last Theorem (FLT) -} When $n>2$,
the equation $x^n + y^n = z^n$ has no nontrivial integer solutions.

\bigskip\noindent After reading the explanations in the previous
section, a skeptic asked the author what the difference is between
the previous argument that the Collatz $3n+1$ Conjecture is
unprovable and the following argument that Fermat's Last Theorem is
unprovable (which cannot possibly be valid, since Fermat's Last
Theorem was proven by Wiles and Taylor in the last decade of the
twentieth century \cite{b:Wei05a}):

\bigskip\noindent \textit{Invalid Proof that FLT is unprovable:}
Suppose that we have a computer program which computes $x^n+y^n-z^n$
for each $x,y,z \in \mathbb{Z}$ and $n>2$ until it finds a
nontrivial $(x,y,z,n)$ such that $x^n+y^n-z^n=0$ and then halts.
Obviously, Fermat's Last Theorem is equivalent to the assertion that
such a computer program can never halt. In order to be certain that
such a computer program will never halt, it is necessary to compute
$x^n+y^n-z^n$ for each $x,y,z \in \mathbb{Z}$ and $n>2$ to determine
that $x^n+y^n-z^n \neq 0$ for each nontrivial $(x,y,z,n)$. Since
this would take an infinite amount of time, Fermat's Last Theorem is
unprovable. \qed

\bigskip\noindent This proof is invalid, because the assertion that
``it is necessary to compute $x^n+y^n-z^n$ for each $x,y,z \in
\mathbb{Z}$ and $n>2$ to determine that $x^n+y^n-z^n \neq 0$ for
each nontrivial $(x,y,z,n)$" is false. In order to determine that an
equation is false, it is not necessary to compute both sides of the
equation - for instance, it is possible to know that the equation
$6x+9y=74$ has no integer solutions without evaluating $6x+9y$ for
every $x,y \in \mathbb{Z}$, since one can see that if there were any
integer solutions, the left-hand-side of the equation would be
divisible by three but the right-hand-side would not be divisible by
three.

\bigskip\noindent \textbf{Question -} So why can't we apply this same
reasoning to show that the proof that the Collatz $3n+1$ Conjecture
is unprovable is invalid? Just as it is not necessary to compute
$x^n+y^n-z^n$ in order to determine that $x^n+y^n-z^n \neq 0$, is it
not possible that one can determine that the Collatz algorithm will
converge to one without knowing what the algorithm does at each
iteration?

\bigskip\noindent \textbf{Answer -} Because what the Collatz
algorithm does at each iteration \underline{is} what determines
whether or not the Collatz sequence converges to one \cite{b:Fei05},
it is necessary to know what the Collatz algorithm does at each
iteration in order to determine that the Collatz sequence converges
to one. Because the exact values of $x^n+y^n-z^n$ are
\underline{not} relevant to knowing that $x^n+y^n-z^n \neq 0$ for
each nontrivial $(x,y,z,n)$, it is not necessary to compute each
$x^n+y^n-z^n$ in order to determine that $x^n+y^n-z^n \neq 0$ for
each nontrivial $(x,y,z,n)$.

\bigskip\noindent \textbf{Exercise -} You are given a deck of $n$
cards labeled $1,2,3,...,n$. You shuffle the deck. Then you perform
the following ``reverse-card-shuffling" procedure: Look at the top
card labeled $k$. If $k=1$, then stop. Otherwise, reverse the order
of the first $k$ cards in the deck. Then look at the top card again
and repeat the same procedure. For example, if $n=7$ and the deck
were in order $5732416$ (where $5$ is the top card), then you would
obtain $4237516 \rightarrow 7324516 \rightarrow 6154237 \rightarrow$
$3245167 \rightarrow 4235167 \rightarrow$ $5324167 \rightarrow
1423567$. Now, we present two problems:

\begin{itemize}
\item Prove that such a procedure will always halt for any $n$ and
any shuffling of the $n$ cards.
\item Find a closed formula for the maximum number of iterations that
it may take for such a procedure to halt given the number of cards
in the deck, or prove that no such formula exists. (The maximum
number of iterations for $n=1,2,3,...,16$ are
0,1,2,4,7,10,16,22,30,38,51,65,80,101,113,139 \cite{b:Slo05}.)
\end{itemize}

\noindent It is easy to use the principle of mathematical induction
to solve the first problem. As for the second problem, it turns out
that there is no closed formula; in other words, in order to find
the maximum number of iterations that it may take for such a
procedure to halt given the number of cards $n$ in the deck, it is
necessary to perform the reverse-card-shuffling procedure on every
possible permutation of $1,2,3,...,n$. This property of the
Reverse-Card-Shuffling Problem in which there is no way to determine
the outcome of the reverse-card-shuffling procedure without actually
performing the procedure itself is called \textit{computational
irreducibility} \cite{b:Wol02}. Notice that the notion of
computational irreducibility also applies to the Collatz $3n+1$
Conjecture and the Riemann Hypothesis in that an infinite number of
irreducible computations are necessary to prove these two
conjectures.

Stephen Wolfram, who coined the phrase ``computational
irreducibility", argues in his famous book, \textit{A New Kind of
Science} \cite{b:Wol02}, that our universe is computationally
irreducible, i.e., the universe is so complex that there is no
general method for determining the outcome of a natural event
without either observing the event itself or simulating the event on
a computer. The dream of science is to be able to make accurate
predictions about our natural world; in a computationally
irreducible universe, such a dream is only possible for very simple
phenomena or for events which can be accurately simulated on a
computer.

\section{Open Problems in Mathematics}

In the present year of 2006, the most famous unsolved number theory
problem is to prove the following:

\bigskip\noindent \textbf{Goldbach's Conjecture -} Every even number
greater than two is the sum of two prime numbers.

\bigskip\noindent Just like the Collatz $3n+1$ Conjecture and the
Riemann Hypothesis, there are heuristic probabilistic arguments
which support Goldbach's Conjecture, and Goldbach's Conjecture has
been verified by computers for a large number of even numbers
\cite{b:Guy04}. The closest anyone has come to proving Goldbach's
Conjecture is a proof of the following:

\bigskip\noindent \textbf{Chen's Theorem -} Every sufficiently large
even integer is either the sum of two prime numbers or the sum of a
prime number and the product of two prime numbers \cite{b:Che66}.

\bigskip\noindent Although the author cannot prove it, he believes the
following:

\bigskip\noindent \textbf{Conjecture 1 -} Goldbach's Conjecture is
unprovable.

\bigskip\noindent Another famous conjecture which is usually mentioned
along with Goldbach's Conjecture in mathematics literature is the
following:

\bigskip\noindent \textbf{The Twin Primes Conjecture -} There are
infinitely many prime numbers $p$ for which $p+2$ is also prime
\cite{b:Guy04}.

\bigskip\noindent Just as with Goldbach's Conjecture, the author cannot
prove it, but he believes the following:

\bigskip\noindent \textbf{Conjecture 2 -} The Twin Primes Conjecture is
undecidable, i.e., it is impossible to know whether the Twin Primes
Conjecture is true or false.

\section{Conclusion}
The $P \neq NP$ problem, the Collatz $3n+1$ Conjecture, and the
Riemann Hypothesis demonstrate to us that as finite human beings, we
are all severely limited in our ability to solve abstract problems
and to understand our universe. The author
hopes that this observation will help us all to better appreciate the
fact that there are still so many things which G-d allows us
to understand.

\section*{Acknowledgments}

I thank G-d, my parents, my wife, and my children for their support.

}
\centerline{\rule{0pt}{19pt}\rule{72pt}{0.4pt}}

\end{document}